# Fractal seismicity and seismic risk

## G.M. Molchan

*Institute of Earthquake Prediction Theory and Mathematical Geophysics, Russian Academy of Science, Moscow, Russia*

*E-mail*: molchan@mitp.ru

**Abstract.** Recently, attempts have been made to take into account the fractal properties of seismicity when mapping the long-term rate of earthquakes. The paper touches upon the theoretical aspects of fractality and provides a critical analysis of its applications to the problems of seismic risk

## 1. Introduction.

The long-term rate of seismic events is a basic parameter of seismicity for seismic zoning and seismic risk problems. That parameter is based on the Gutenberg-Richter law, which defines the log-rate of magnitude $\geq M$ events in a domain $\Delta$ as the following linear relation:

$$\log \lambda(\Delta, M) = a - b(M - M_0) , \quad M_- \leq M \leq M_+ \tag{1}$$

There is as yet no commonly accepted methodology for mapping the parameters $(a,b)$ that can incorporate the magnitude range $\Delta M = (M_-, M_+)$. However, broadly speaking, the conventional approach is based on seismotectonic regionalization of an area of study, the choice of suitable zones that must have constant values of each of the above parameters, and subsequent statistical estimation of these parameters (see, e.g., [Molchan and Podgaetskaya, 1973; Molchan et al., 1996]).

Since damaging earthquakes are few, one can call in question both the elements subject to zoning and the choice of the magnitude ranges.

In this connection it is of interest to discuss the series of publications (Kossobokov and Nekrasova, 2004; Nekrasova and Kossobokov, 2006, 2009; Nekrasova et al., 2011; Nekrasova et al., 2015; Kossobokov and Soloviev, 2018] to be referred to as [K+] in what follows. These publications stress the importance of taking "the fractal nature of earthquakes" into account. The authors considered the problem of seismic risk for mega-cities around the world to conclude that the neglect of fractality in the distribution of seismicity appreciably underestimates the risks. "The overall level of recurrence underestimation is too large to be neglected in calculations of seismic risk and losses, which are so badly needed in order to take measures for disaster prevention and for mitigating the impact".

The authors propose an alternative approach that can be translated from Russian in two ways: General Law of Earthquake Similarity ( GLES) or Unified Earthquake Scaling Law (USLE). The second version has been used by other writers and has fundamentally different implications (see below). Therefore we will associate the term GLES with the papers [K+]. The proposed approach predicts the rate of magnitude $\geq M$ events in a domain $\Delta_L$ of size $L$, $\Delta_L \subset \Delta_{L_0}$, according to the relation

$$\lambda(\Delta_L, M) = A 10^{-B(M-M_0)} (L/L_0)^C \quad , \quad l_0 \leq L \leq L_0 \quad (2)$$

Here, the target area $\Delta_L$ is located at the center of the enclosing domain $\Delta_{L_0}$; $A$ is the expected annual rate of $\geq M_0$ events in $\Delta_{L_0}$, $C$ is the fractal dimension of $\geq M$ earthquake epicenters in $\Delta_{L_0}$.

The first occurrence of (2) in seismological literature dates back to Keilis-Borok et al. (1989) as an example of seismicity similarity.

The rest of this paper is organized as follows: in Section 2, we shall describe the GLES method and its applications to seismic risk problems; in Section 3, we remove contradictions by treating (2) as the normalization in the Bak seismicity laws. Later on, in Section 4, we discuss the optimal choice of the parameter $C$ using rigorously formulated concepts of fractality and multifractality. The reading of our material is facilitated by summaries appended to sections 3 and 4.

## 2. Seismicity models: the conventional models and the GLES.

For comparison purposes let us start from one of seismicity model based on the traditional law of recurrence [Molchan et al., 1997, 1999]. The model incorporates relation (1) in the form of a piecewise linear relation over space and magnitude. When the magnitude range $\Delta M$ is fixed, the activity parameter $a$ varies over area more than does $b$. The linear size of the area with a constant value of $b$ (a $b$ zone) is noticeably greater than the maximum source length possible for the magnitude range $\Delta M$. The higher the magnitudes, the larger are the $b$ zone. This is due to the necessity of excluding possible artefacts in the form of "characteristic events" [Wesnousky, 1994], expanding thereby the domain where model (1) is applicable.

The result is to have a multiscale representation of (1) with parameters $(a, b, \Delta M)_i$ for suitable areas. We now illustrate this by an example. Molchan et al. (1997, 1999) used two magnitude levels ($\Delta_1 M = 3.5 - 5$ and $\Delta_2 M = 5 - 7$) to describe potentially damaging earthquakes in Italy ($M > 3.8$). The parameter $a$ of the first level is constant at elements of seismotectonic regionalization ($a$ zones) that have typical sizes of 40–130 km in length and 20–30 km in width. Further diminution of these areas is hampered by the available earthquake statistics. Zones that have constant values of $b$ are composed of seismically connected $a$ zones and the $b$ zone size is in agreement with the magnitude ranges $\Delta_i M$, i.e., $b$ zones need to be tectonically uniform in the proper scale. Such $b$ zones of the first level were 10, while those of the second level were 3. The space distribution of the parameter $a$ in the $b$ zone of the second level depends on the nontrivial problem of spatial location of large earthquakes. The relevant data are few; hence the solution is not unique and may depend on concrete applications for deriving upper/lower bounds for risk estimates.

Note the following elements in the above model that are important for comparison with the GLES method: (1) an informal approach to areal mapping of the parameters in the frequency–magnitude relation, and (2) a differentiated (over magnitude and area) incorporation of conditions that would retain similarity in the recurrence of events. The real obstacles to similarity are examined in detail by Ben-Zion (2008).

In the example we considered, the chief contribution into the number of model parameters is due to elements of the original seismotectonic regionalization and, when the amount of earthquake data is remembered, to *a* zones of the first level.

*The GLES methodology*. The parameters $(A, B, C)$ in (2) are mapped at nodes {s} of a standard grid with a step $l_0$. Each set of the three parameters at an element is determined by the seismicity observed in a $L_0 \times L_0$ cell centered at s. To obtain these observations, we covered the $L_0 \times L_0$ cell with a standard grid of step $2^k l_0$, where $k = 0,1,......$ is the scale level of the grid. For each cell of an admissible level $k$, $\Delta^{(k)} \subset L_0 \times L_0$, we found the number of events with magnitudes $\geq M$ for the period $T$, $n(\Delta^{(k)} | M, T)$. The quantities $n(\Delta^{(k)} | M, T) / T$ were treated as estimates of $\lambda(\Delta_L, M)$ with $L = 2^k l_0$ in (2). The entire set of these estimates for all admissible levels $k = 0,1,...$ was used to find the parameters $(A, B, C)$. To enhance reliability of the estimates, the original grid was rotated and the parameters averaged. The standard procedure involved four binary levels with $l_0 = (1/4)^0 \approx 28 km$ and $L_0 = 2^0 \approx 222 km$; the original sizes are given in degrees of the terrestrial meridian.

We see that the smallest areas used in the above approaches are comparable in size, being $l_0 \approx 30 km$. For this reason the difficulties involved in estimating the rate of events at this scale are identical. The traditional approaches overcome these difficulties by enlarging the original regionalization elements based on seismotectonic arguments, while the GLES method, where the smallest cells are defined by a formal grid, uses the similarity hypothesis in the form of (2). The seismicity rate of events in the minimal cell $l_0 \times l_0$ is estimated, as we have seen, by immersing this cell in the set of similar to it from $L_0 \times L_0$. This enlargement of the data set is to enhance reliability for the estimation of seismicity rate in $l_0 \times l_0$ based on (2). This expectation is quite justified, provided the hypothesis (2) holds and the cell $l_0 \times l_0$ of interest is *typical* among similar subcells of $L_0 \times L_0$.

The italicized term can be explained by the following example. Suppose all seismicity concentrates at a narrow fault at a distance greater than $l_0$ from the original node of the grid. In that case the original square $l_0 \times l_0$ cannot be typical on the scale $l_0$, although this does not in the least affect the estimation of $A, B, C$. This can be seen as follows. The parameters $A$ and $B$ are independent of epicenter geometry, while the parameter $C$ must reflect the dimension of the fault which does not intersect the considered subcell $l_0 \times l_0$. Consequently, the seismicity rate in this subcell will be overestimated by the GLES method.

It is generally not so obvious to identify atypical cells. However, their presence is manifested on a large scale. Indeed, the smallest subcells from $L_0 \times L_0$ are absolutely equivalent in the estimation of the GLES parameters. The authors refer the parameters $A, B, C$ to the center of the area $L_0 \times L_0$ just for keeping on the safe side, since they regard the fractal properties of seismicity inhomogeneous over area, [K+]. For this reason the estimate of seismicity rate for the central subcell $l_0 \times l_0$ would be logical to extend to the other cells in theoretical analysis. In that case the rate of $\geq M_0$ events in $L_0 \times L_0$ can be found by summing the rates over all smallest subcells. Since the number of such cells is $(L_0/l_0)^2$, and bearing in mind that the fractional dimension $C$ does not exceed the dimension of the space, $C \leq d = 2$, we obtain the result

$$A = A(l_0/L_0)^C \times (L_0/l_0)^2 = A(L_0/l_0)^{2-C} \geq A \tag{3}$$

The relation is contradictory, unless when $C = d = 2$. To exclude the contradiction (3), we have to conclude that the condition $C < 2$ implies existence atypical subcells in $L_0 \times L_0$.

The foregoing argument reveals a substantial difference between models (1) and (2). Our reasoning above implicitly used the additive property which the measure of seismicity rate $\lambda(\Delta, M)$ ought to possess. That means that any nonintersecting subsets $\Delta_i$ satisfy the following equality:

$$\lambda(\cup\Delta_i, M) = \sum_i \lambda(\Delta_i, M). \qquad (4)$$

Relation (2), being a model for the measure of seismicity rate, must possess this property, but the parameterization involved here prevents this. Exceptions may include the cases of integer values of $C$: 1, 2 or 3. That fact can be proved similarly to (3), provided one uses the correct space dimension, i.e., d=2 is replaced with d= 1 or 3. For any $C$, the GLES model retains additivity only for the sell of the standard grid at intervals of $L_0$, because for these, (2) becomes identical with the ordinary frequency–magnitude relation.

This loss of additivity in (2) appreciably limits the area of its applicability. So far, all examples of application that we are aware of were concerned with the assessment of seismic risk for cities. Here we quote one of these from [Kossobokov and Soloviev, 2018]:

"estimating the frequency of earthquakes for Petropavlovsk-Kamchatsky whose area is $S_{PK} = 400 km^2$, where $A = 0.12$, $B = 0.86$, $C = 1.26$ as obtained from seismicity data for all of Kamchatka whose area is $S_{Kam} = 270000 km^2$, we get the result that the underestimate of earthquake frequency within the city limits when normalized by area can amount to

$$\nu = (S_{Kam}/S_{PK})/(S_{Kam}/S_{PK})^{C/2} = 675^{0.37} > 11 \text{ times''}. \qquad (5)$$

The above calculation requires some explanation. Keeping to the standard of the authors, we shall consider $L_0 \times L_0$ cell of 1 degree of the terrestrial meridian with area $S_0 = 111.2^2 km^2$. Proceeding in accordance with a model that assumes a uniform distribution of epicenters in an $L_0 \times L_0$ cell, we find the rate of $\geq M_0$ events in the area $S_{PK}$: $\lambda_0 = A(S_{PK}/S_0)$. A similar estimate derived in accordance with (2) must be $\lambda = A(\sqrt{S_{PK}/S_0})^C$. The parameters $A$ are identical, because they give the rate of $\geq M_0$ events in an $L_0 \times L_0$ sell. The result is $\nu = \lambda/\lambda_0 = (S_0/S_{ПК})^{1-C/2} = 3.56$ instead of 11.1 or 7.6, which is an estimate from another paper [Nekrasova and Kossobokov, 2009].

The estimate $v = 11.1$ is based on the comparison of the GLES model with the null hypothesis that assumes a uniform seismicity distribution over all of Kamchatka. When no further specifications concerning the area have been made, the hypothesis looks extremely doubtful, since the mapped activity parameter varies by some tens of times near Petropavlovsk-Kamchatsky (in the W–E direction) [Gusev et al., 1980]. For this reason estimates of efficiency for the GLES can be regarded with some confidence, only if (1) the seismicity in the area of study is typical of the entire territory where the GLES parameters were determined, and (2) the hypothesis of a *uniform* distribution of epicenters in the territory under study is reliable. In other words, a correct comparison between the two methods should incorporate real alternatives related to seismic regionalization. These are much more complicated than the hypothesis of a uniform seismicity distribution, even in standard $1^0 \times 1^0$ sells.

One study in risk estimation for cities with a million plus population in seismic zones is (Keilis-Borok et al., 1985). Although the 1980 data were crude enough, the prediction of intensity $I \geq 8$ shaking for cities based on traditional approaches has proved quite reliable for monitoring periods of 10 and 20 years in the study referred to. The conclusion is related to large groups of cities with similar-sized populations. Under these conditions, the probabilistic law of large numbers provides stability of seismic risk assessments.

### 3. **The GLES as a normalizing factor.**

The fact that model (2) is not additive means that the measure of seismicity rate is more complicated. The model can be corrected, if we consider the history of the problem.

The term GLES is synonymous with the term "Unified Earthquake Scaling Law" (USLE). However, the meaning of GLES is radically different from the alternative construct, which was first introduced by Bak et al. (2002). To clarify the meaning of the USLE, let us consider a seismic zone G and superimpose on it a regular grid at intervals of $L$. Let $\xi(L \times L)$ be a statistic of seismic events of magnitude $\geq M$ in a $L \times L$ area acquired during a time $\Delta T$. Following and generalizing Bak, we shall say that seismicity has the USLE property, if there is such a normalization $a_L = f(L, M, \Delta T)$ of statistics $\xi(L \times L)$ that the distribution of $a_L \xi(L \times L)$ averaged over all cells of size $L$ is independent of the parameters $L, M, \Delta T$. The averaged distribution can be interpreted as the distribution of the normalized statistic in an area that has been selected in a *random* manner. One can have a stronger variant of the USLE, in which the normalized statistic has the same distribution for all seismic cells. The term *Unified* must then be replaced with *Universal*, according to [Corral, 2003].

Bak et al. (2002) and Corral (2003, 2004) considered an example of a USLE statistic in the form of the time interval $\tau_L$ between successive events in a $L \times L$ cell. The statistic is of interest because its natural normalization must be proportional to the rate $\lambda(\Delta_L, M)$ of events of magnitude $\geq M$ in $\Delta_L = L \times L$, because the mean $E\lambda(\Delta_L, M)\tau_L = 1$ is independent of the parameters, when the seismicity is stationary. Bearing in mind the Gutenberg-Richter law over magnitude and the fact that the distribution over space is fractal (for more detail see below), Bak et al. (2002) proposed the normalizing factor for $\tau_L$ in the form

$$a_L(\tau_L) = A \cdot 10^{-B(M-M_0)} \cdot L^{d_f} \qquad (6)$$

where $d_f$ is a fractal dimension of the epicenters.

Considered formally, the right-hand sides of (6) and (2) are identical parametrically, but are different as to the meaning and the goals they are intended to achieve. In addition, if we proceed on the basis of the estimation methods proposed by the authors, the fractal parameters $d_f$ and $C$ have different types of dimensionality (see below).

According to Bak and Corral, the normalized distribution of $\tau_L$ provides a fairly good fit to the Gamma distribution whose density is $p(x) = cx^{\gamma-1}e^{-x}$. This is true for full catalogs with aftershocks. Theoretical analysis showed [Molchan, 2005] that $\tau_L$ as a USLE statistic occurs only for a homogeneous Poisson flow of events. This is not the case for real seismicity and the value of the parameter $\gamma$ which is significantly different from 1.

The USLE property of $\tau_L$ is well visualized due to a graphical representation of distributions $a_L \tau_L$ in log-log scales [Molchan and Kronrod, 2007]. This representation clearly shows that the distribution tails are consistent among themselves, but it masks the deviations around moderate values of these statistics. The $a_L \tau_L$ distribution behavior near small values is power-type due to the Omori law, while at large values is exponential due to the poissonian property of main events. Thus, the USLE law for $\tau_L$ reflects the already known empirical regularities.

Staying within the Bak approach, it would be a more natural procedure to look for the optimal normalization of $\xi(L \times L)$, making the distribution of $a_L \xi(L \times L)$ in a *random* $L \times L$ cell to be *extremely weakly* dependent on $\Delta T, L, M$, while remaining *undegenerated*.

The phrase *'extremely weakly'* means the choice of a normalization such that the distributions of $\{a_L \xi(L \times L)\}$ with different $L$ are extremely close to one another in an appropriate metric. Since the comparison of the distributions due to Bak and Corral were somewhat deficient, the deviations between distributions were estimated by Molchan and Kronrod [2005] in the Levy metric [Feller, 1966]. This metric measures the greatest difference between the plots of two distributions in the (-1,1) direction. The Levy metric is the most sensitive instrument to detect deviations between distributions around the central values of the relevant random variables.

The fact of being *non-degenerated* in application to a limiting distribution means that the normalized statistic is finite and does not vanish when $L \ll 1$. Our attention to small scales is due to the fact that it is only for $L \to 0$ that one can make comparatively correct statements about fractality.

The problem of optimal normalization has been discussed by Molchan and Kronrod (2005, 2007) for the two statistics $\tau_L$ and $n_L$, where $n_L$ is the number of $\geq M$ events in a *randomly selected* $L \times L$ cell for time $\Delta T$. When it is impossible to distinguish between typical and atypical cells, the analysis of long-term $n_L$ statistics in a random $L \times L$ square is the natural alternative to relation (2). One purely theoretical example of seismicity in which $n_L$ has the exact USLE property was proposed by Pisarenko and Golubeva [1999].

Since $En_L = \lambda(\Delta_L, M)\Delta T$, the normalization of $n_L$ is again related to the rate of events in $L \times L$:

$$a_L(n_L) = [\Delta T \cdot A 10^{-B(M-M_0)} \cdot (L/L_0)^{d_f}]^{-1}. \qquad (7)$$

Relations (6,7) determine the normalization structure alone, but do not imply coincidence between the parameters involved in both of the statistics $\tau_L, n_L$. The optimal value of $d_f$ can depend both on the choice of the particular statistic and on our definition of *randomness* or, which amounts to the same thing, the weight of the $L \times L$ sell in the averaging of the distributions.

We shall select the $L \times L$ cell with probability

$$w_p(L \times L) = \lambda^p(L \times L)/\Lambda_p, \qquad (8)$$

where $\Lambda_p$ is constrained by $\sum w_p(L \times L) = 1$. The option $p = 0$ assumes all seismogenic cells to be equally probable. The option is the one used Bak et al. (2002). The option $p = 1$ looks the most natural, since the weight of a cell is proportional to the number of events observed in it. Large values of $p > 1$ can locate places of high seismicity to varying degrees of resolution.

Molchan and Kronrod (2005) considered a space-homogeneous multifractal measure of seismicity (to be exactly defined in what follows) to develop a constructive description of non-uniqueness in the choice of the optimal parameter $d_f$ and to confirm their derivations using the seismicity on the San Andreas fault as an example. Among other things, they found that the optimal normalization of $n_L$ for M $>$ 2 events is reached with $d_f \approx 1.8$ for the case of the weight parameter $p = 0$ and with $d_f \approx 1.2$ for the case $p = 1$. These empirical estimates of $d_f$ are appreciably different and both apply to the scale range $L = 10 – 100$ km. Nekrasova and Kossobokov [2009] used $C \approx 1.2$ to estimate the effect of the GLES method for Los Angeles by the value of (5): $v = 6.2$. When $C \approx 1.8$, the effect is reduced by nearly a factor of four: $v \approx 1.6$, which emphasizes the fact that the fractal parameter should be chosen in a correct manner.

*Summary*. The model of seismicity rate (2) does not have the additive property, and therefore has to be corrected. Treating the GLES as a normalization of statistics $n_L$ in the randomized cell, one can appreciably reduce (but not remove) its distribution dependence on scale. As a result, the adjusted relation (2) takes the stochastic form:

$$\lambda(\Delta_L, M) = A 10^{-B(M-M_0)} (L/L_0)^C \xi(\Delta_L), \quad L << L_0, \tag{9}$$

where all one knows about the $\{\xi(\Delta_L)\}$ is merely that the distribution of each $L$-set of the $\{\xi(\Delta_L)\}$ is weakly dependent on the scale. Here, the cells $\Delta_L$ make grid of scale $L$, while $L_0$ is the scale of the area that has homogeneous (see below) fractal properties of the seismicity measure. The corrected model (9) can no longer provide point estimates of seismicity rate, and at best can be used to construct confidence intervals for $\lambda(\Delta_L, M)$, which reduces its practical value.

The model (9) was tested using events (M>2) on the San Andreas fault [Molchan and Kronrod, 2005], and obviously has to be validated before being applied to the particular case of a region and a magnitude range. Some nontrivial examples of distributions of $\{\xi(\Delta_L)\}$ can be found in [Molchan and Kronrod, 2005].

The choice of $C$ is not unambiguous and calls for a special analysis to which we pass.

## 4. Multifractality and the parameter C.

An understanding of the parameters C, $d_f$ in (2, 6) can be reached through accurate definitions of fractality and multifractality. These concepts are intimately related to infinitesimal spatial scales that are unobtainable by analyses of seismicity. However, there is no other way to understand those pitfalls which are hidden when using these concepts formally at the macrolevel.

The rate of $\geq M$ events in $\Delta$ cell, $\lambda(\Delta, M)$, determines a measure $\mu(\Delta)$ on subsets of the seismic zone, when $M$ is fixed. Treating the measure $\mu(\Delta)$ as a purely mathematical object, it is referred to as a *multifractal*, if, roughly speaking, its support can be decomposed into a sum of subsets $S_\alpha$ such that, taking a suitable set of vicinities $\Delta_L$, which is proper to each particular point $g \in S_\alpha$, the measure shows a type $\alpha$ singularity:

$$\log \mu(\Delta_L) \sim \alpha \log L \quad , \quad L << 1 \,. \tag{10}$$

(The notation $x \sim y$ means that $x/y \approx 1$ when $L << 1$).
Note that a smooth measure on a plane generally has the parameter $\alpha = 2$.

The whole set of $(\alpha, f(\alpha))$ pairs, where $f(\alpha)$ is the fractional dimension of $S_\alpha$, defines the multifractal spectrum of the measure. Theoretical studies associate fractional dimensions with the Hausdorff dimension and with box dimension in physical, less rigorous, applications. By definition, the spectrum of a monofractal (or simply a fractal) consists of a single pair, while the spectrum of a *homogeneous multifractal* is the same for different parts of the measure support.

The spectrum of a homogeneous multifractal contains a point at which the singularity is identical with the dimension, $\alpha_0 = f(\alpha_0)$. This kind of dimension is called *information dimension* $D_1$, because all points having this kind of singularity make a set of complete measure: $\mu(S_{\alpha_0}) = \mu(\cup S_\alpha)$. In other words, (10) is valid with $\alpha = D_1$ in a suitable small vicinity of any *typical* point for the measure $\mu(\cdot)$. The concept of being typical has been discussed at the macrolevel in Section 2. As to the microlevel, the antipodes of typical points in a multifractal measure are diverse, and they are composed of point sets $S_\alpha$ of homogeneous singularity $\alpha \neq D_1$. Similarly to the macrolevel situation, the sets $S_\alpha$ are unknown. Any singularity (10) can generally be detected by statistical techniques, but with no indication of the location.

One constructive criterion for a measure to be multifractal is the approximately linear relation

$$\log \sum \mu^p (L \times L) \sim \tau(p) \log L, \quad L \ll 1 \tag{11}$$

for small cell scales and different fixed values of $p$. For simplicity of notation, the measure has been normalized here, i.e., it has a full unit mass, so that $\tau(1) = 0$. The summation in (11) is over cells of non-zero $\mu$-measure, e.g., over seismogenic cells.

It is known that, if the curve $[-\tau(p)]$ is convex, the values of $\tau(p)$ completely specify the multifractal spectrum of the measure. In particular, the set of all singularities like (10) in the regular situation is identical with the set of values of $d\tau(p)/dp := \dot{\tau}(p)$. We remind the exact relations for the main (in applications) fractional dimensions: *box* dimension of the support of the measure $D_0 = -\tau(0)$, *information* dimension $D_1 = \dot{\tau}(1)$, and *correlation* dimension $D_2 = \tau(2)$ [Hentschel & Procaccia, 1983; Feder, 1988].

Both the dimensions and the singularities of a measure provide a formal explanation of the power-law parameterization in (2,6). On the one hand, (10) is equivalent to the crude relation $\mu(\Delta_L) \propto L^\alpha$ for small scales near points from $S_\alpha$. (The notation $x \propto y$ means that $x/y$ varies slowly as $L \to 0$). Any set $S_\alpha$ of a homogeneous multifractal measure is everywhere dense on its support. For this reason any singularity of the measure can be viewed as a candidate for the fractal parameter in (2,6). Among these candidates the $\alpha = D_1$ singularity is to be preferred, since it is associated with typical points of the measure.

Consider the fractal dimension $f(\alpha)$. The number of cover elements $\Delta_L$ for an arbitrary fractal set $S$ increases as a power law function: $N(L) \propto L^{-d_S}$, where $d_S$ is the box dimension of $S$. Take the measure support as $S$. In that case the average number of cover elements would be $\mu(\Delta_L) \propto N(L)^{-1} \propto L^{D_0}$. For this reason the parameter $d_f$ in (2,6) can well be assumed to be the dimension $D_0$, and this has been done by Bak et al. [2002]. In a similar manner, assuming $S = S_\alpha$, we get $\mu(\Delta_L) \propto L^{d_f}$ with $d_f = f(\alpha)$. Both of these interpretations of the power law behavior of the measure are consistent, when $\alpha = f(\alpha) = D_1$, i.e., when $S$ is the set of typical points of the measure.

The possibilities for the fractal parameter in (2,6) described above are still insufficient to make the optimal choice of it, when we deal with a random cell $\Delta_L$. According to [Molchan and Kronrod, 2005], when a randomized cell is described by the distribution (8) with $p \geq 0$, then

$$d_f^{opt}(p) = \tau(p+1) - \tau(p) \quad . \tag{12}$$

Now because $\tau(1) = 0$, it follows that $d_f^{opt}$ is identical with the box dimension, $d_f^{opt}(0) = D_0$, when p=0, and with the correlation dimension, $d_f^{opt}(1) = D_2$, when p=1. These empirical estimates of $d_f^{opt}$ for the San Andreas Fault have confirmed this theoretical conclusion at scales of 10-100 km.

The choice of a random cell with a frequency that is proportional to the number of events in the cell looks natural. It is intimately connected with information dimension. For this reason the theoretical prediction of $d_f^{opt}(1) = D_2$ instead of $D_1$ is, on the face of it, unexpected. As a matter of fact, however, (12) results from a balanced incorporation of typical and atypical points in a multifractal measure, depending on the procedure employed for random choice, hence is a kind of averaging applied to dimensions.

A method of estimation for the parameter $C$ in [K+] leads to a correlation dimension $D_2$, but smoothed by rotating the grid. The choice $D_2 \neq 2$ says in favors of the presence of atypical cells, while (2) itself excludes this possibility.

*The range of scale.* In practice, the range of scale $L$ in (2,6) is bounded away from large and small values, i.e.,

$$\delta_- < \lg L < \delta_+ \quad . \tag{13}$$

One extra requirement

$$\delta_+ - \delta_- \approx 1 \div 2 \tag{14}$$

is typical of discussions on *nontriviality of similarity* in physics literature [Malkai et al., 1997]. Under conditions (11), (14) with stable estimates of $\tau(p)$ at some parameters $p$, we can say that the physical object in question looks like a fractal or a multifractal ($\tau(p) \neq cp$) in the range of scale (13).

The above requirements, in particular, the stability and the condition (14), have proved rather stringent on the world and regional catalogs available at the time of [Molchan and Kronrod, 2009] work. We analyzed world seismicity to identify only six regions (southern California, M>2; Kamchatka, M>3.5; New Zealand, M>2.5; Central American arc, M>4; Costa Rica, M>3.2; Greece, M>3 ; Garm (Central Asia), M>1.7) where positive conclusions could be reached for some individual areas to have multifractal seismicity in the following log range of scale: $\delta_+ - \delta_- \in [1, 1.6]$. Unfortunately, rigorous analyses of fractality in seismicity [see e.g. Goltz, 1997; Harte, 2001; Molchan and Kronrod, 2009] are still few.

The works [K+] recommend the following fixed range of log-scale in applications: $\delta_+ - \delta_- \approx \lg 8 = 0.9$, which is below 1. Estimates of $C$ are not always reliable under these conditions; a formal averaging of the estimates derived by rotating the standard $L_0$ grid makes the estimates stable, while leaving the question of their reliability open.

*Summary*. The seismicity rate measure for small events can be considered to be a multifractal in some individual, very few so far, regions worldwide. For multifractal seismicity relation (12) defines the optimal fractal parameter $C$ in (2). The parameter serves the measure of seismicity rate in a random cell of size $L \ll L_0$, where $L_0$ is the scale of multifractal homogeneity, while randomness is described by the probabilities given by (8). The parameter $C$ is identical with correlation dimension, when the probability of a cell choice is proportional to seismic activity in the cell.

The model (2) postulates that fractal dimension is independent of magnitude, which is not at all obvious, when supports of large and small events are compared. The former usually tend to occur on faults, while the latter are dispersed over area, which generally inflates the fractal dimension Under condition (14) a reliable analysis of fractality for large events seems doubtful.

## 6. Conclusions

We have examined a method for mapping long-term seismicity rate, called the *General Law of Earthquake Similarity,* GLES . The method is based on the meaningful hypothesis, not as yet used in applications, stating that the measure of seismicity rate is fractal. Practical applications of the method encounter serious difficulties: (1) fractality / multifractality for the measure of seismicity rate has been reliably confirmed in few regions of the world, and for small earthquakes only. Any extrapolation of the fractal properties of low magnitude seismicity to high magnitudes, which is important because damaging earthquakes are few, requires a serious justification; (2) the GLES measure of seismicity rate does not possess the necessary property of additivity, which leads to contradictions and the following unjustified inference: "One consequence from the fractal nature of earthquakes and, in particular, their distribution over area, is the *commonly occurring underestimation* of traditional estimates of earthquake hazard", [K+]. For this reason the use of fractal properties of seismicity in seismic risk assessment remains an open question.

(3) One should not exaggerate the role of fractality for risk assessment, where large events are dominant, while the justification of even their spatial support is difficult. The crux of the matter is that the damaging effect of a magnitude M event is felt in a spatial zone $G_{J(M)}$ (e.g., within a shaking zone of level J). For this reason the risk for a point facility $g_0$ is determined by the total rate of M events in zone $G_{J(M)}$ centered at $g_0$ [Keilis-Borok et al., 1984]. In that case it is sufficient to have a smoothed and therefore not fractal measure of seismicity plus control integrals of seismic rate in suitable areas.

Examples of correct use of GLES are related to the Bak methodology, where the basic relation (2) is interpreted as a suitable normalization of seismicity at different scales (see Section 3). In this way it is possible to obtain nontrivial distribution laws that are weakly dependent on the scale. However, these do not characterize the seismicity in specific cells of a fixed size, but the seismicity in a cell that has been selected randomly among these cells. The idea of a randomized cell is certainly of interest for obtaining laws of similarity in seismicity

**Acknowledgments.** My colleagues V. Pisarenko, M. Rodkin, and L. Romashkova have read the preliminary version of this paper. Their informal remarks have prompted a radical revision of the text, and I am sincerely grateful to them.

### References

. *Bak,P.,Christensen, K.,Danon, L.,Scanlon,T*. Unified scaling law for earthquakes //Phys.   Rev.Lett., 2002, 88, 178501.

*Ben-Zion, Y*. Collective behavior of earthquakes and faults: continuum discrete  transitions, progressive evolutionary changes and different dynamic regimes// *Rev. Geophys.,*2008, 46, RG4006, p. 1–70.

*Corral A*. Local distribution and rate fluctuation in unified scaling law for earthquakes // Phys. Rev.E., 2003, 68, 035102(R).

*Corral A*. Universal local versus united global scaling laws in the   statistics of seismicity// Physica A, 2004, 340 , p. 590 – 597.

*Feder  J*. Fractals, Plenum Press, New York, 1988, 283p

*Feller W*. An introduction to probability theory and it applications, V.2, New York, 1966,750 p.

*Goltz  C*. Fractal and Chaotic Properties of Earthquakes, Springer, Berlin, 1997, 182p.

*Gusev A. Kondratenko, A., Potapova, O., Fedotov, S., and Shumilina, L.*, Kamchatka and the Commander Islands, in *The Seismic Regionalization of the USSR* (in Russian), Nauka, 1980, pp. 269-283,

*Harte D*. Multifractals: Theory and Applications*,* Boca Raton, Chapman & Hall, 2001, 248 p

*Hentschel, H. G. E., & Procaccia, I*. The infinite number of generalized dimensions of fractals and strange attractors// Physica D, 1983, 8, 435–444.

*Keilis-Borok V.I., Kronrod, T.L., and Molchan, G.M.*, Earthquake risk of the world's largest cities (intensity VIII shaking), in *Vychislitel'naya seismologiya* (Computational Seismology), vol. 16, *Mathematical Modeling and Interpretation of Geophysical Data,* Allerton Press, 1985, pp. 92-113.

*Keilis-Borok, V., Kossobokov, V., and Mazhkenov, S.*, On similarity in the spatial distribution of seismicity, in *Theory and Algorithms for Interpretation of Geophysical Data* (in Russian), Keilis-Borok, V.I., Ed., Moscow: Nauka, 1989, pp. 28-41 (*Vuchislitelnaya Seismologiya* 22).

*Kossobokov, V. and Nekrasova, A.*, A general law of earthquake similarity: A global map of parameters, in *The Analysis of Geodynamic and Seismic Processes* (in Russian), Keilis-Borok, V.I., Ed., Moscow: Nauka, 2004, pp. 160-176 (*Vychislitelnaya Seismologiya* 35)

*Kossobokov, V. and Soloviev, A.*, Pattern recognition in earthquake hazard assessments, *Chebushevsky Sbornik,* 2018, vol. 19, no. 4, pp. 53-89.

*Malkai O., Lidar D., Biham O., Avnir D*. Scaling range and cutoffs in empirical fractals// Phys.Rev. E., 1997, 56, p. 2817-2828.

. *Molchan G*. Interevent time distribution of seismicity: a theoretical approach //Pure. appl. Geophys., 2005, vol.162, p.1135-1150.

*Molchan G., Kronrod T.* On the spatial scaling of seismicity rate//Geophys. J. Int , 2005, vol.162, p.899-909.

*Molchan G., Kronrod T*. Seismic interevent time: a spatial scaling and multifractality // Pure appl. geophys., 2007, vol.164, p.75-96.

*Molchan G., Kronrod T*. The fractal description of seismicity// Geophys. J. Int., 2009, vol.179, N3, p.1787-1799.

*Molchan G., Kronrod T. and Panza G*. Multi-scale seismicity mode for seismic risk// Bull. of the Seismological Society of America, 1997, vol.87, N5, p.1220-1229

*Molchan, G. and Podgaetskaya, V.,* The parameters of global seismicity, in *Computational and Statistical Methods for Interpretation of Seismic Data* (in Russian), Keilis-Borok, V.I., Ed., Moscow: Nauka, 1973, pp. 44-66 (*Vychislitelnaya Seismologiya* 6).

*Molchan, G.M., Kronrod, T.L., Dmitrieva, O.E., and Nekrasova, A.K.,* Hazard-oriented multiscale seismicity model: Italy, in *Computational Seismology and Geodynamics,* vol. 4, AGU, 1999, pp. 138-156.

*Nekrasova, A. and Kossobokov,* V., A general law of earthquake similarity: The area around Lake Baikal, *Doklady Akad. Nauk (Geofizika),* 2006, vol. 407, no. 5, pp. 679-681

*Nekrasova, A. and Kossobokov, V.*, A general law of earthquake similarity: Megacities and urbal agglomerations, in *Some Problems of Geodynamics* (in Russian), Keilis-Borok, V.I., Ed., Krasnodar, 2009, pp. 265-300 (*Vychislitelnaya Seismologiya* 39).

*Nekrasova,A.,Kossobokov, V.,Peresan,A., Aoudia A.,Panza G.* A multiscale application of the unified scaling law for earthquakes in Central Mediterranean area and Alpine region // Pure appl.Geophys.,2011, vol.168, p.207-327**.**

*Nekrasova A.,Kossobokov V., Parvez I., Tao X.* Seismic hazard and risk assessment based on the unified scaling law for earthquakes// Acta Geod. Geophys. **,** 2015, vol.50, p.21-37**.**

*Pisarenko, V. and Golubeva, T.*, The use of stable laws in seismicity models, in *Computational Seismology and Geodynamics*, AGU, 1999, vol. 4, pp. 127-137 (*Vychislitelnaya Seismologiya* 28).

*Wesnousky S.* The Gutenberg-Richter or characteristic earthquake distribution. Which is it?//Bull.Seismol.Soc.Amer, 1994, vol.84, P.1940-1959.